\documentclass[preprint,14pt]{elsarticle}

\usepackage{amsmath, amssymb, latexsym, amscd, amsthm,amsfonts,amstext}
\usepackage[mathscr]{eucal}
\usepackage{graphicx, subfig}
\usepackage[]{algorithm2e}
\usepackage[english]{babel}
\usepackage[colorinlistoftodos]{todonotes}

\newtheorem{remark}{Remark}[section]
\numberwithin{equation}{section}

\newcommand{\tbs}{\setminus}

\newcommand{\JJ}{\mathbf{J}}
\newcommand{\xx}{\mathbf{x}}

\newcommand{\nn}{\mathbf{n}}

\newcommand{\dO}{\partial\Omega}

\graphicspath{
{figures/}
}

\journal{NeuroImage}

\begin{document}
\begin{frontmatter}

\title{FEM-based Scalp-to-Cortex data mapping via the solution of the Cauchy problem}

\author[affil1]{Nikolay Koshev\corref{cor1}}
\author[affil1,affil2]{Nikolay Yavich}
\author[affil1,affil2]{Mikhail Malovichko}
\author[affil1]{Ekaterina Skidchenko}
\author[affil1]{Maxim Fedorov}

\cortext[cor1]{Corresponding author. Email: n.koshev@skoltech.ru}

\address[affil1] {CDISE, Skolkovo Institute of Science and Technology}
\address[affil2] {Applied computational geophysics lab, Moscow Institute of Physics and Technology}

\begin{abstract}
	We propose an approach and the numerical algorithm for pre-processing of the electroencephalography (EEG) data, enabling to generate an accurate mapping of the potential from the measurement area - scalp - to the brain surface. The algorithm based on the solution of ill-posed Cauchy problem for the Laplace's equation using tetrahedral finite elements linear approximation. Application of the proposed algorithm sufficiently increases the spatial resolution of the EEG technique, making it comparable with much more complicated intracranial EEG techniques. 
\end{abstract}


\end{frontmatter}

\section{Introduction} \label{sec:intro}

The majority of existing methods for EEG data processing and source analysis falls into the following categories: parametric inversions (the dipole fitting), current-reconstruction methods (\cite{PM1994,PM2002,Hamalainen1994,Lin2006,Baillet2001,Grech2008}) and, somewhat less frequently, beamforming methods (\cite{VanVeen1997,Mosher1998,Hillebrand2005, Sekihara2008}. The approaches are very different (sometimes, fundamentally). All of them, however, use the EEG measurements as enter data and, thus, highly depend on its quality. 

The measured electrical potential is strongly distorted by outer (with respect to the brain) head tissues. 
The current research is aimed to improve the EEG data via partial excluding the influence of the outer tissues on the EEG signal. 

Since the outer compartments (tissues) do not contain any sources producing the objective signal, the electric potential related to the brain sources satisfies the Laplace's equation there. 
Measuring the potential on the part of the head surface makes it possible to state the Cauchy problem for the Laplace's equation in order to map the data from the scalp to the brain surface. Such mapping is an auxiliary problem, allowing to sufficiently increase the accuracy of source localization due to better quality and spatial resolution of enter data. 

The Cauchy problem for the data mapping in the context of EEG was considered in \cite{clerc2007}, where authors propose to use the boundary elements method (BEM) in order to construct the cost functional for further minimization of it. The method, however,  probably may have some drawbacks common for the BEM techniques. It is known that the conventional BEM (i.e., double layer formulation) accuracy degrade when the distance between the source and one of the surfaces becomes smaller \cite{DeMunck}. Beyond this, when two surfaces approach each other, the resulting system of linear equations tends to be singular. Further, conventional BEM cannot incorporate conductivity anisotropies. Finally, numerical comparison presented in \cite{Hyde} indicated much higher modelling errors in the BEM versus finite element method (FEM). 

In the current paper, we use the technique based on the mixed quasi-reversibility (MQR) method for linear finite elements and proposed by L.Bourgeois in \cite{Bourgeois}. 
The method allows to reduce the Cauchy problem to the system of linear equations per compartment, built using two regularization parameters, the choice of which is also considered in \cite{Bourgeois, Bourgeois_2005}. 
As we show later, the application of our algorithm enables us to 'focus' a measured signal, sufficiently improving its spatial resolution even in complicated cases of in-brain source currents. 

Regarding practical use of the proposed method, it can be applied either directly to detect active parts of the cortex or as an intermediate procedure to propagate data from scalp to cortex before employing other processing techniques. Though in this study we focused only on mathematical and numerical aspects, we believe that the proposed method may serve as a basis for fast and accurate algorithms for analysis of real EEG data.

The paper is organized as follows. In Section \ref{sec:equations}, we describe the conductivity model of the human head, discuss the approximations being used and write out the governing equation of the EEG. In Section \ref{sec:Cauchy}, we discuss the Laplace's equation for the sourceless compartments of the head and present some important notes about our implementation of the MQR method. The Section \ref{sec:numerical_results} presents some numerical results. Concluding remarks are given in Section 5. 

\section{Mathematical description of EEG} \label{sec:equations}
The current section is aimed to introduce the notations, which will be used in paper, and remind the reader about some aspects of EEG together with its assumptions and equations. 

\subsection{The computational domain and sources} \label{sec:domains}

The head volume in general and the brain in particular consist of many parts and organs: scalp, skull bone, cerebrospinal fluid, the brain itself and its components. 
Using the macroscopic approximation, we assume the conductivity to be identical across the particular compartment of the head, i.e., we expect the conductivity to be the piecewise-constant distribution over the computational domain. 

\begin{figure}[h!]
	\begin{center}
		\includegraphics[width=0.5\linewidth]{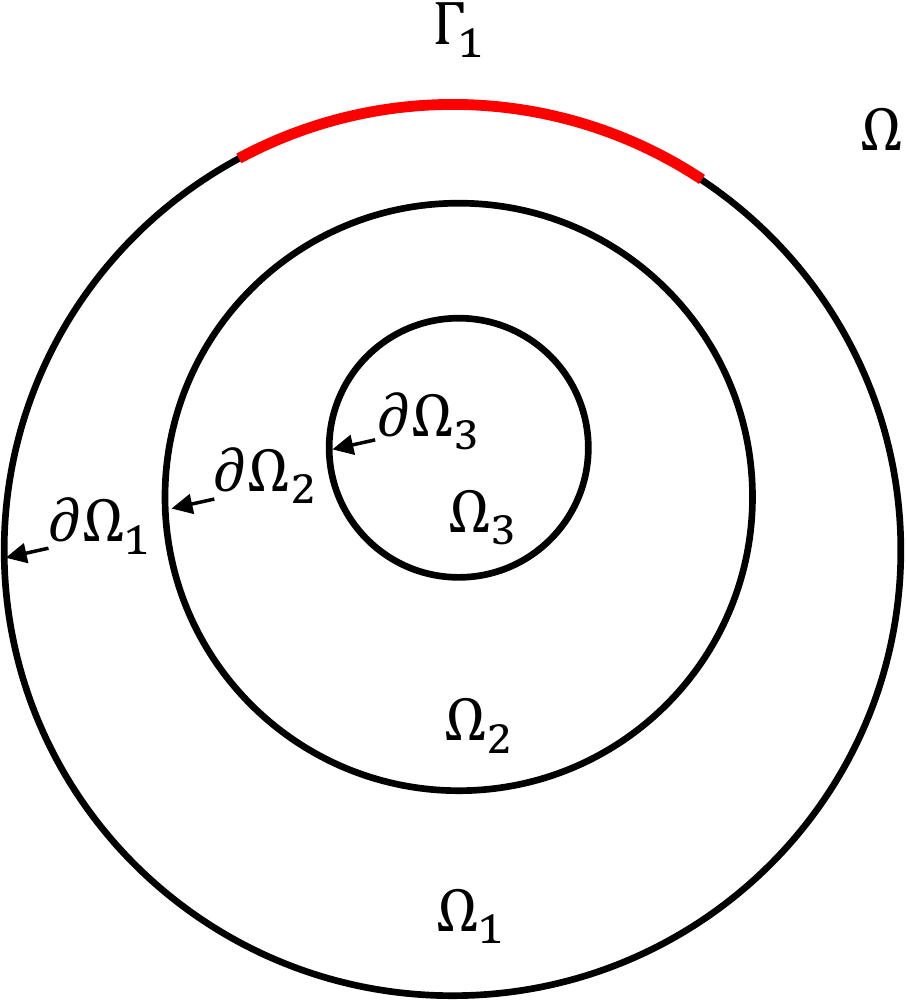}
	\end{center}
	\caption{Computational domains and boundaries.
	\label{fig:domain_structure}
	}
\end{figure}

Consider the domain $\Omega\subset \mathbb{R}^3$ with a piecewise-smooth 
boundary $\dO$. The domain under consideration represents the volume of the head and is assumed to consist of several subdomains 
\begin{equation*}
	\Omega_i, i=1...N_d: \Omega = \bigcup_{i=1}^{N_d} \Omega_i,
\end{equation*}
representing corresponding $N_d$ organs inside the head (see Fig.\ref{fig:domain_structure}). Each organ expected to have the constant electrical conductivity $\sigma_i$ (see \cite{tissue_conductivity1}).

\textbf{We further assume a nested domain topology of the introduced partitioning.}
We denote the outer boundary of each volume $\Omega_m$ as $\dO_m$. For convenience, indexing can be done starting from outer areas (scalp, skull) to inner areas. In such way, the outer subdomain $\Omega_1$, for example, is bounded with the surfaces $\dO_1$ (outer surface) and $\dO_2$ (inner surface). The outer head surface in such indexing is $\dO \equiv \dO_1$. 

In the EEG technique, the measurements are being performed in a bounded area of the surface located commonly at the top of a head. In further consideration, this part of the boundary, which we call \emph{accessible part of the boundary}, is represented by the subdomain of the outer surface $\Gamma_1\subset\dO_1$ (see Fig. \ref{fig:domain_structure}). Generally, if the domain $\Omega$ consists of more than two sourceless domains, we define the area $\Gamma_i$ for each of them as follows: 
\begin{equation*}
	\Gamma_i = \begin{cases}
		\Gamma_1, & i=1; \\
		\dO_i, & i\ge2.
	\end{cases}
\end{equation*}

For further discussion we also need to introduce the rest of the boundary $\Pi_i$ for the subdomain $\Omega_i$, which we call \emph{inaccessible part of the boundary}. This part of the boundary contains unknown data, which we will need to restore. 
\begin{equation}
	\Pi_i = \begin{cases}
		\big(\dO_1 \tbs \Gamma_1\big) \cup \dO_2, & i=1; \\
		\dO_i, & i\ge2.
	\end{cases}
\end{equation}

EEG method measures the electrical potential evoked by the biochemical currents inside a head. There are two main kinds of such currents: currents generated in the brain and the muscle currents. Since the EEG technique aimed to study the electrical activity of the brain, the signal of interest is generated by brain currents. Muscle currents have much higher frequencies and thus can be easily filtered. Therefore, we can assume the current sources are located only inside the brain (more accurately - on the brain surface), while all other areas do not contain electrical sources: $supp\big(\JJ(\xx)\big) \subset \Omega_{N_d}$.

The structure of the calculation domain for simplified head models is depicted in Fig.\ref{fig:domain_structure}.

\begin{remark}\label{remark:domains}
	For simplicity, in the current article, we take into account only simple head models consisting of two domains: the brain is represented by the inner domain $\Omega_2$, and the outer tissues (scalp, skull, cerebrospinal fluid, etc.) are described by the only one outer domain $\Omega_1$. Despite that, all other reasoning is correct with respect to numerous sourceless domains, and, thus, the algorithm described in further consideration can be easily applied to a real head model containing a massive amount of sourceless domains. 
\end{remark}

\subsection{Governing Equation}

As mentioned above, the conductivity $\sigma(\xx), x\in \Omega$ assumed to be a piecewise-constant distribution over the volume under consideration (head). The conductivity of the media outside the head expected to be zero: 
		\begin{equation}
			\sigma(\xx) = \begin{cases}
				\sigma_i, & \xx\in \Omega_i \subset \Omega; \\
				0, & \xx\notin \Omega.
			\end{cases}
		\end{equation}
		
Given the frequencies of the neural currents within the interval of 1-100 Hz, the quasi-static approximation is justified. This was proved in numerous well-known works on EEG, e.g. \cite{Sarvas, DeMunck, Plonsey_1969, Hamalainen, Zakharova}. 

Let the volumetric distribution of a primary neuronal current density (the source) be denoted as $\JJ^p$. Under the assumptions, the electrical potential satisfies the equation
\begin{eqnarray}
	\nabla\cdot(\sigma\nabla U) = \nabla\cdot \JJ^p, \quad \xx\in\Omega. \label{U_equation} \
\end{eqnarray}
The potential further satisfies Neumann boundary condition on $\dO$: 
\begin{equation}\label{V_boundary_cond}
	\nn\cdot\nabla U = 0, \quad x\in \partial\Omega.
\end{equation}
This boundary condition follows from the assumption of zero conductivity of the media outside the head \cite{DeMunck}.

We assume the existence of the described in subsection \ref{sec:domains} \emph{accessible part of the boundary} $\Gamma_1 \subset \Omega$, on which the electrical potential $U(\xx)$ is known: 
\begin{equation}\label{boundary_cond_V_B}
	U|_{\Gamma_1} = u(\xx),  \quad \Gamma_1 \subset 
		\partial\Omega \subset \mathbb{R}^3.
\end{equation}

\begin{remark}\label{remark:conductivity}
Equation (\ref{U_equation}) contains the coefficient $\sigma(\xx)$ - distribution of the tissues conductivity over a head volume. This distribution can be obtained using methods such as magnetic resonance imaging (MRI) or computed tomography (CT). There are also other techniques for conductivity distribution acquisition. Additionally, some standard models of such distribution are often employed. Regardless of the approach, in the current work we assume the distribution $\sigma(\xx)$ to be known at every point $\xx\in\Omega$. 
In biomedical applications, the discussed conductivity varies within a considerable interval. For example, the conductivity of the brain is about one order greater than the conductivity of the tissues located outside the brain (see, e.g., \cite{tissue_conductivity1, tissue_conductivity2}). This fact causes distortion of the electrical potential measured on the outer surface of the domain (head). This situation will be discussed later, in Section \ref{sec:numerical_results}. In the current paper, the conductivity of the inner part of the model, representing the brain, was taken as a constant equal to $2.2$ S/m, and the conductivity of the outer sourceless part was taken as $0.2$ S/m. 
\end{remark}

\section{The Cauchy problem for sourceless domains} \label{sec:Cauchy}

Areas outside a brain volume consist of skin, skull bones, muscles. Muscles in a tensed state can also produce rather strong currents and, consequently, electric field. However, they generate currents with much higher frequencies, which allows to filter it out. Accordingly, we consider the "useful" signal to be produced only by brain currents. 
Thus, in our consideration we can expect the tissues outside a brain do not contain any sources, which assumption is important for further reasoning. 

In sourceless homogeneous volumes the equation (\ref{U_equation}) will take a form:
\begin{eqnarray}
	\Delta U = 0, & \xx\in \dO_i, & i=1,...,N_{sl},\label{laplace_U}\
\end{eqnarray}
where $N_{sl}$ - the number of sourceless domains. Indeed, the equation above can be obtained from (\ref{U_equation}) by nulling the right-hand side due to absence of sources with the assumption that the conductivity $\sigma_i = const$. 

We can also write the Dirichlet and Neumann boundary conditions for sourceless domains: 
\begin{eqnarray}
	U(\xx)|_{\Gamma_i} = u_i(\xx), \\ 
	\nn\nabla U(\xx)|_{\Gamma_i} = \frac{\sigma_{i-1}}{\sigma_{i}} \nn\nabla \widetilde{U}|_{\Gamma_{i}} \equiv g_i(\xx); 
\end{eqnarray}
where $\widetilde{U}$ denotes the potential on the outer side of the interface $\dO_i$. The boundary conditions are obtained from requirements for potential and its normal derivative to be continuous inside the computational domain $\Omega$. Assuming the conductivity $\sigma(\xx) = 0, \xx\notin\Omega$, we can easily obtain zero Neumann condition, common for EEG problem. 

\subsection{Statement of the problem} \label{sec:statement}

The equation (\ref{laplace_U}) enable us to state an auxiliary problem, which we named the \emph{propagation problem}. 

\textbf{Propagation problem for the electrical potential}. \emph{Assume the function $u(\xx), \quad \xx\in\Gamma_1$ to be known. Find the function $U(\xx)$, such that:}
\begin{eqnarray}\label{Cauchy_U}
	\Delta U = 0, & \xx \in \Omega_i; \label{extra1}\\
	U|_{\Gamma_1} = u_i(\xx), & \xx \in \Gamma_i; \label{extra2}\\
	\nn\cdot\nabla U|_{\Gamma_i} = g_i(\xx), & \xx\in \Gamma_i \label{extra3}.
\end{eqnarray}

\begin{remark}\label{remark:medical applications}
In medical applications, the interface between the brain and cerebrospinal fluid contains sources. Moreover, the mentioned interface is the only area, where the electrical currents are possible except muscle currents. Due to this fact, from the mathematics' point of view, the Cauchy problem for the Laplace's equation can be stated in this case only for open domain, which does not contain the brain surface itself. Despite this, our numerical experiments showed that its solution (i.e., backpropagation of the potential to the brain surface) is accurate enough and can be used in practice. This caused by the fact we do not compute the potential only on boundary, but compute it in the whole sourceless domain. Thus, we can estimate the electrical potential at the points arbitrary closed to the desirable surface. The respective numerical experiments will be shown in Section \ref{sec:numerical_results}.
\end{remark}

The Cauchy problem, in general, is a well-known problem, and the Cauchy problem for the Laplace's equation with incomplete boundary data is one of the classical examples of an ill-posed problem. During the last years, a number of validated approaches to its solution were presented. The approaches can be divided into three main categories: approaches based on the optimization (least square, Tiknonov's regularization etc), methods based on construction via quasi-reversebility of the linear systems of equations and solving it and iterative approaches (see, for example, \cite{Engl}). 

The first category includes classical Tikhonov's regularization \cite{Berntsson, Li}, which is well-established and mostly used, but have some issues with the solution accuracy \cite{JIIP-2016}. Additionally, it includes a set of powerful methods proposed by the group of M. Klibanov together with one of the authors of the current paper. The approach based on the construction of strictly convex Tikhonov-like functional weighted with a problem-depending Carleman weight functions \cite{JIIP-2016, NONRWA-2016}. Despite a great accuracy of the solution, the method, however, is rather hard to implement and is better suitable for nonlinear problems, for which it was designed. Other solutions based on the optimization approaches are presented, for example, in \cite{clerc2007}.

The second category is probably the most known category of the approaches. It based on usage of quasi-reversibility method, which is being used in order to state a weak formulation of a problem, which, after discretization, is turned into linear system of algebraic equations. Some methods are presented, for example, in \cite{Bourgeois, Bourgeois_2005, Cao, Cao_Klib_Per_2009, Darde-2013}. 

\subsection{The method} \label{sec:method}

The equations (\ref{extra1}-\ref{extra3}) are the Cauchy problems for the Laplace's equation. Uniqueness of this problem was proved in many ways. The most effective and beautiful proof can be found in \cite{LavRomShish}. Solution of the problem considered in a set of works (see, for example, \cite{Bourgeois, Engl, Berntsson, Li, NONRWA-2016, Cao, Cao_Klib_Per_2009}).

In the current paper, we use the MQR method provided for linear finite-element approximations in \cite{Bourgeois} and \cite{Bourgeois2006}. As one can see below, the method depends on two regularization parameters. The most efficient way of choosing the regularization parameters is the balancing principle, which is described in \cite{Cao_Klib_Per_2009}. 

For simplicity, we will next consider only the case of two compartments: 'inactive' sourceless compartment $\Omega_1$ and 'active' subdomain $\Omega_2$ containing the electrical sources. Consider the Cauchy problem (\ref{Cauchy_U}). In our consideration the part of the boundary $\Gamma_1$ is shown, e.g., in Fig.\ref{fig:domain_structure}. 

We introduce the following notations:
\begin{eqnarray}\label{MQR_subspaces}
	V_0 = \{h\in H^1(\Omega_1) : h|_{\Gamma_1} = 0 \}, \label{V0}\\
	V_1 = \{h\in H^1(\Omega_1) : h|_{\Pi_1} = 0 \}, \label{V1}\\
	\tilde{V}_0 = \{h\in H^2(\Omega_1) : h|_{\Gamma_1} = u(x)\}. \label{V00}
\end{eqnarray}

Let $0<\epsilon\ll1, 0<\delta\ll1$ be two fixed small numbers (regularization parameters). 

\textbf{Weak formulation of MQR method \cite{Bourgeois}}. \emph{Find a pair of functions $(U, \lambda) \in \tilde{V}_0 \times V_1$, such that:}
\begin{eqnarray}\label{weak_QR}
	\epsilon \int\limits_{\Omega_1} \nabla U\cdot \nabla h dx + \epsilon \int\limits_{\Omega_1} uhdx + 
		\int\limits_{\Omega_1} \nabla h \cdot \nabla \lambda dx = 0, & \forall h\in V_0; \label{weak_qr1}\\
	\int\limits_{\Omega_1} \nabla U \cdot\nabla \mu dx - \delta\int\limits_{\Omega_1} 
		\nabla\lambda\cdot\nabla\mu dx - (1+\delta)\int\limits_{\Omega_1} \lambda\mu dx = 
		\int\limits_{\Gamma_1} g_1 \mu d\Gamma_1, & \forall \mu \in V_1. \label{weak_qr2}
\end{eqnarray}

In \cite{Bourgeois} one can find proofs of existence and uniqueness of the solution, its stability and convergence of the solution of (\ref{weak_QR}) and the exact solution of the problem (\ref{Cauchy_U}) with $(\epsilon, \delta) \to 0$, when the error in the enter data $u(\xx) \to 0$.

\paragraph{The finite element approximation} 
In order to make further consideration shorter we start with the fact the second term of the equation (\ref{weak_qr1}) and the third term in (\ref{weak_qr2}) can be omitted, the justification of which one can find in \cite{Bourgeois}. We also note that since we consider the Cauchy problem for the outer domain $\Omega_1$, the function $g(\xx) \equiv 0$. 

Omitting the index under the function $u_1(\xx)$, we can rewrite the system under consideration as follows: 

\begin{eqnarray}\label{final_system}
	\epsilon \int\limits_{\Omega_1} \nabla U\cdot \nabla h dx + 
		\int\limits_{\Omega_1} \nabla h \cdot \nabla \lambda dx = 0, & \forall h\in V_0, \label{weak_qr1_2}\\
	\int\limits_{\Omega_1} \nabla U \cdot\nabla \mu dx - \delta\int\limits_{\Omega_1} 
		\nabla\lambda\cdot\nabla\mu dx = 0, & \forall \mu \in V_1. \label{weak_qr2_2}
\end{eqnarray}
where the function $u(\xx)$ is known on $\Gamma_1$. In our consideration the part of the boundary $\Gamma_1$ is shown, e.g., in Fig.\ref{fig:domain_structure}. 

Among many different methods available to numerical solution of the diffusion equation (finite-difference, finite-volume, see e.g. \cite{Yavich-2007}, finite-element etc.), we used the finite-element method with linear basis functions on tetrahedrons as it combines simplicity and flexibility.

Let $X_h$ be the linear finite element space with the basis $h_i(\xx), i=1,...,N$ (here $N$ is the number of nodes). Denote the following subspaces, which are the finite-dimensional analogs of the corresponding subspaces (\ref{MQR_subspaces}):
\begin{eqnarray}
	X_{0} = \{h\in X_h: h|_{\Gamma_1} = 0\}, \\
	X_1 = \{h\in X_h: h|_{\Pi_1} = 0\}, \\
	\tilde{X}_0 = \{h\in X_h : h - U_0 \in X_0\}.
\end{eqnarray}

Here $U_0$ is the approximation of the function with the following properties:
\begin{eqnarray}
	U_0|_{\Gamma_1} = u(\xx), \quad \partial_n U_0|_{\Gamma_1} = 0, \quad U_0 \in H^1(\Delta, \Omega), 
\end{eqnarray}
where
\begin{equation}
	H^1(\Delta, \Omega) = \{U\in H^1(\Omega): \Delta U \in L^2(\Omega)\}.
\end{equation}

The formulation of the MQR method is still the same as (\ref{weak_qr1_2}-\ref{weak_qr2_2}) with the difference that the pair of functions to be found now belongs to $(U, \lambda) \in \tilde{X}_0 \times X_1$. 

Finite dimensional approximations of the functions $U$, $\lambda$, $g_1$ can be written as follows: 

\begin{eqnarray}\label{femappr_func}
	U(\xx) \approx \sum\limits_{i=0}^N U_i h_i(\xx), & \xx \in \Omega_1, \label{femappr1}\\
	\lambda(\xx) \approx \sum\limits_{i=0}^N \lambda_i h_i(\xx), &  \xx \in \Omega_1, 
	\label{femappr2}
\end{eqnarray}
where $N_{\Gamma_1}$ is the number of nodes belonging to $\Gamma_1$.

For convenience introduce the notation $i(D) = \{i: \xx_i \in D\}$ to be indices of nodes of a finite element mesh, which belong to some domain $D\subset \mathbb{R}^3$. We also use the following notation: 
\begin{equation*}
	A_{gh} = \int\limits_{\Omega} \nabla h_i(\xx) \cdot \nabla h_m(\xx) d\xx, \quad g,h=1,...,N.
\end{equation*}
Substituting (\ref{femappr1}, \ref{femappr2}) into (\ref{weak_qr1}, \ref{weak_qr2}) and assuming $v = h_j \in X_0, \mu = h_k \in X_1$, we obtain:
\begin{eqnarray}
	& \epsilon \sum\limits_{i(\Omega_1)} U_i A_{im} + 
		\sum\limits_{j(\Omega_1)} \lambda_j A_{jm} = 0, 
		& m=m(\Omega_1\tbs\Gamma_1), \label{findim_system1} \\
	& \sum\limits_{l(\Omega_1)} U_l A_{lk} + 
		\delta \sum\limits_{n(\Omega_1)} \lambda_n A_{nk} = 
		b_k, & k=k(\Omega_1\tbs\Pi_1), \label{findim_system2}
\end{eqnarray}
where

\begin{equation*}
	b_k = \sum\limits_{i(\Gamma_1)} u_i \int\limits_{\Gamma_1} \nabla h_k(\xx) \cdot \nabla h_k(\xx) d\xx, \quad k = k(\Omega_1\tbs\Pi_1).
\end{equation*}

Since $U(\xx) \in \tilde{X_0}$, we can state that $U_i = u_i, i\in i(\Gamma_1)$, and the first term of (\ref{findim_system1}) takes the form: 
\begin{equation}
	\epsilon \sum\limits_{i(\Omega_1)} U_i A_{im} = \epsilon \sum\limits_{i(\Omega_1\tbs \Gamma_1)} U_i A_{im} + 
		\epsilon \sum\limits_{i(\Gamma_1)} u_i A_{im}, \quad m \in m(\Omega_1 \tbs \Gamma_1).
\end{equation}
Since the function $\lambda(\xx) \in X_1$, the second term can be written as follows: 
\begin{equation}
	\sum\limits_{j(\Omega_1)} \lambda_j A_{jm} = \sum\limits_{j(\Omega_1 \tbs \Pi_1)} \lambda_j A_{jm}. 
\end{equation}
Reasoning in the same way, we obtain for both equations \ref{findim_system1} and \ref{findim_system2}, one can rewrite these equations as follows: 
\begin{eqnarray}
	& \epsilon \sum\limits_{i(\Omega_1\tbs\Gamma_1)} U_i A_{im} + 
		\sum\limits_{j(\Omega\tbs\Pi_1)} \lambda_j A_{jm} =-a_m, 
		& m=m(\Omega\tbs\Gamma_1), \label{findim_system11} \\
	& \sum\limits_{l(\Omega\tbs\Gamma_1)} U_l A_{lk} + 
		\delta \sum\limits_{n(\Omega\tbs\Pi_1)} \lambda_n A_{nk} = 
		 - b_k, & k=k(\Omega\tbs\Pi_1); \label{findim_system22}
\end{eqnarray}
where 
\begin{eqnarray*}
	a_m = \sum\limits_{i(\Gamma_1)} u_i A_{im}, & m\in m(\Omega_1\tbs\Gamma_1), \\
	b_k = \sum\limits_{i(\Gamma_1)} u_i A_{ik}, & k\in k(\Omega_1\tbs\Pi_1). \\
\end{eqnarray*}

Denote the number of nodes located in some domain $D$ as $N_D$. It is easy to see that in the representation above the discreet system contains $2N_{\Omega} - N_{\dO}$ equations and the same number of variables. It is thus can be solved by an iterative solver. 

\section{Numerical results} \label{sec:numerical_results}
The current section is aimed to present the numerical results on the algorithm described above. We start with two cases of the simple model representing the brain with the inner sphere, and outer tissues with the outer spherical layer. Then, we present the numerical results for more realistic model of the head, based on simplified MRI data (\cite{iso2mesh}). In the current paper, we use simulated data. The simulations were implemented with the finite element numerical solution of the equation (\ref{U_equation}) for the sources/conductivity distributions presented below.

\begin{figure}
	\begin{tabular}{cc}
		\includegraphics[width=0.45\linewidth]{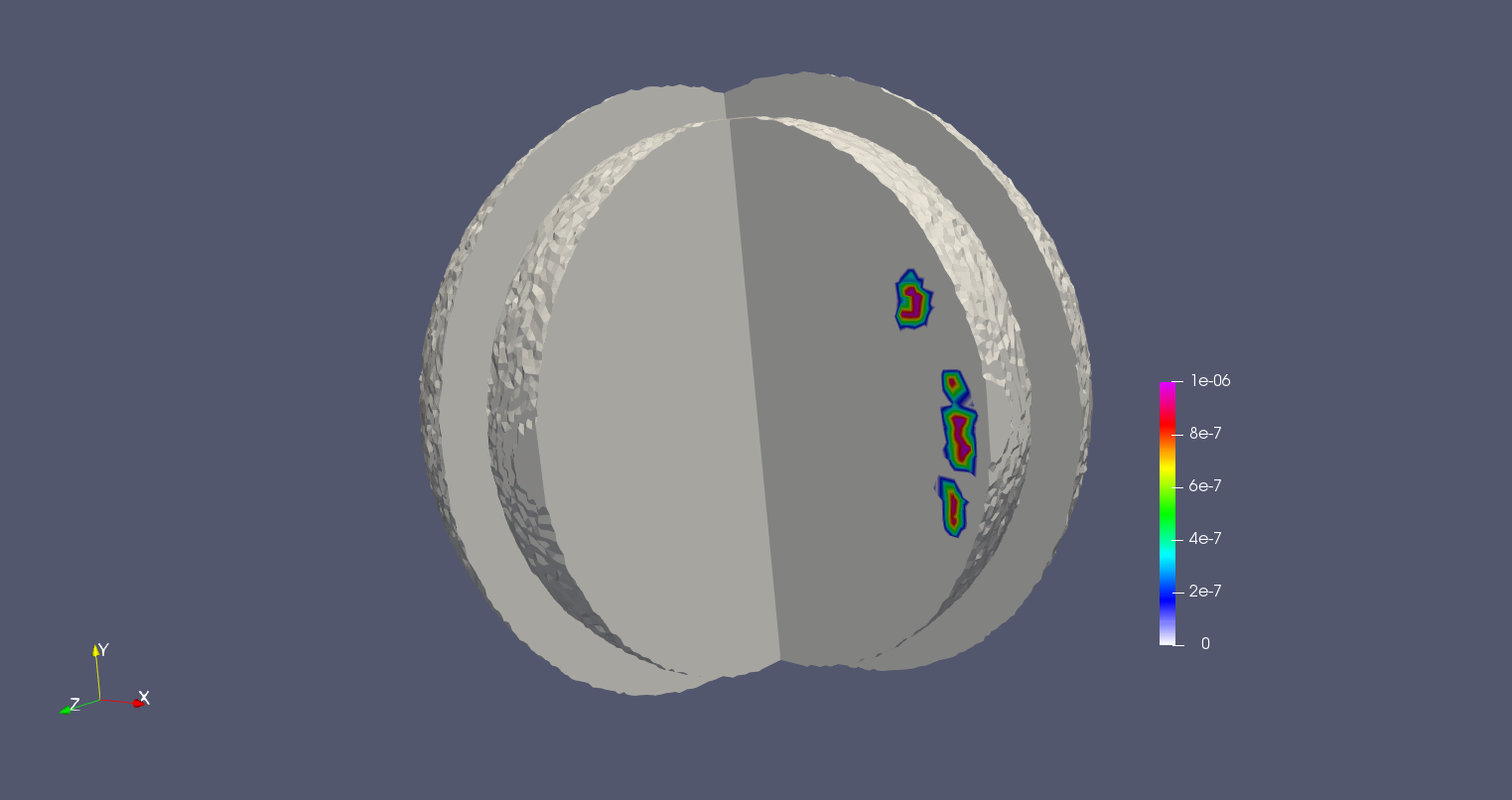} & 
  		\includegraphics[width=0.45\linewidth]{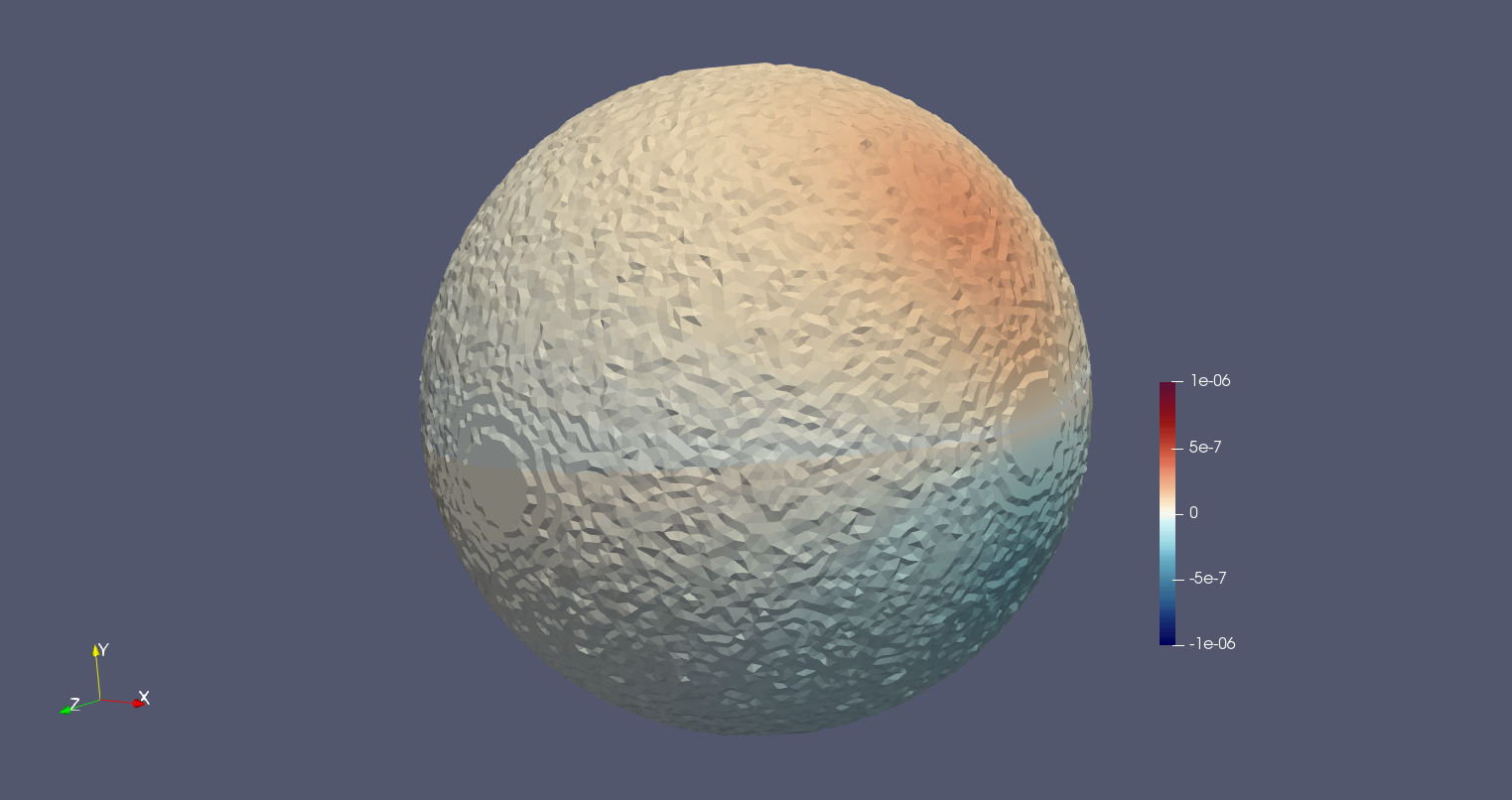} \\ 
		(a) & (b) \\
		\includegraphics[width=0.45\linewidth]{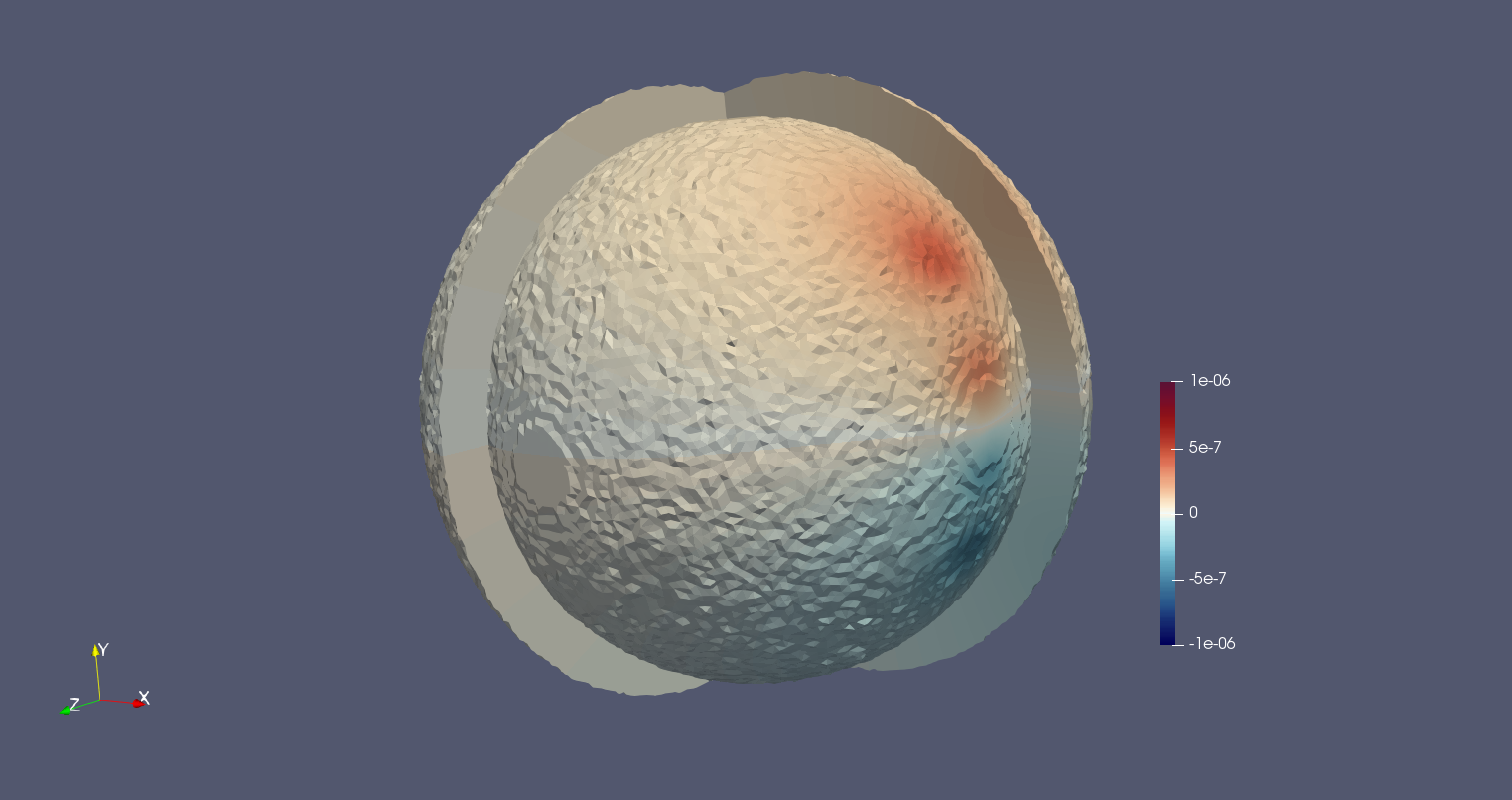} & 
		\includegraphics[width=0.45\linewidth]{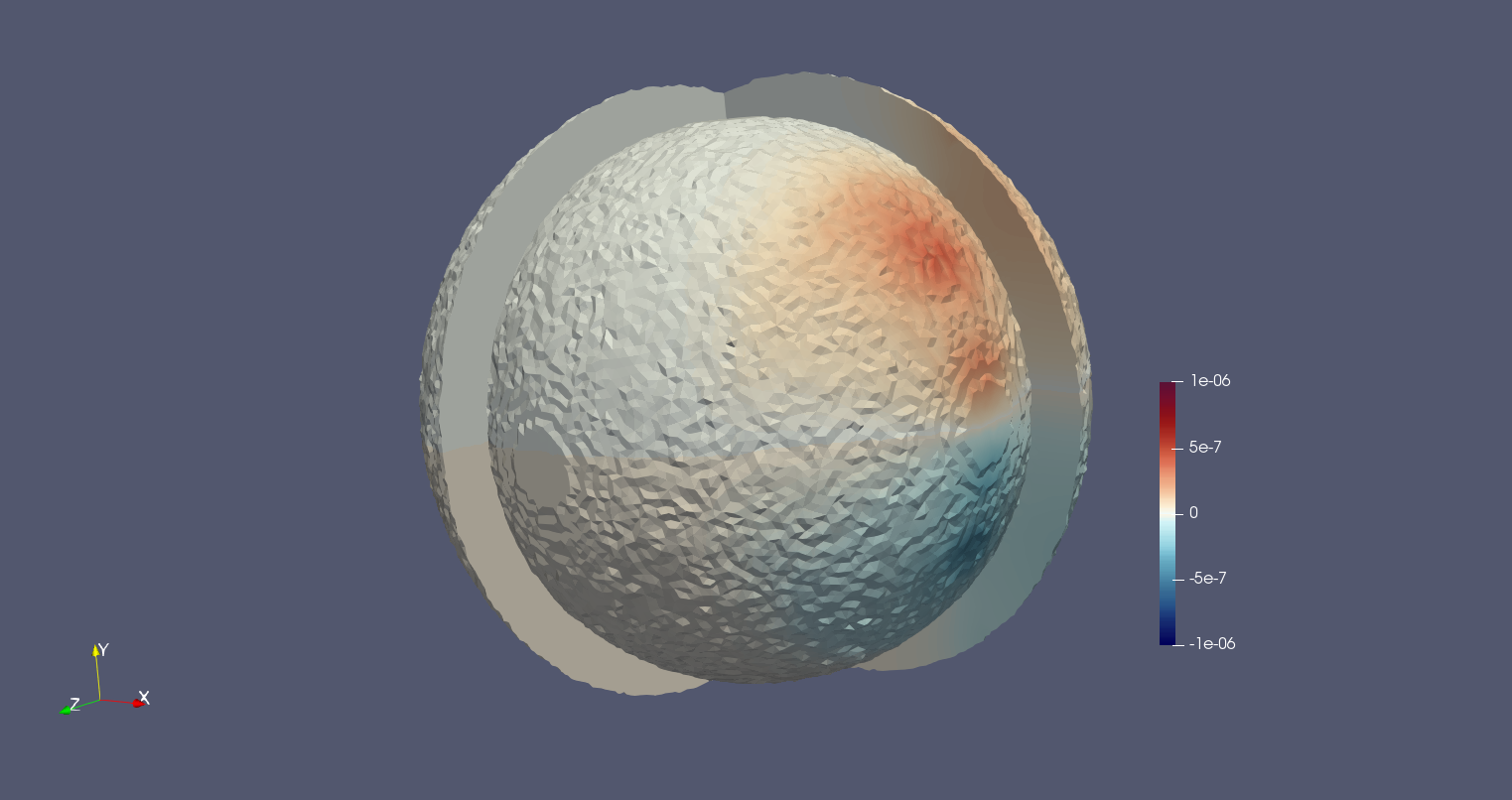} \\
		(c) & (d)
	\end{tabular}
	\caption{Spherical model (sources located under the surface of the active domain): a) simulation of the electrical potential on the outer surface (scalp); b) simulation of the electrical potential on the "active" domain surface; c) simulation of the potential on the surface of the inner sphere; d) the potential on the inner sphere surface mapped from the data depicted in b) via the solution of the Cauchy problem ($\epsilon=0.0001$ and $\delta=0.001$).}
	\label{fig:results_1}
\end{figure}

\subsection{Spherical model with the sources located under the surface of the active domain}
The case considered in this subsection is unrealistic since the currents here are located under the surface of the "active" domain. In truth, the biochemical sources are placed on the interface between a brain and cerebrospinal fluid, i.e., on the brain surface. However, in this case, in the strict sense, the Cauchy problem for the Laplace's equation cannot be stated. Thus, we find it reasonable to start the presentation of our numerical results in the strict case of the sourceless boundary. 

The head here is represented with a simple model, which consists of the spherical "active" domain $\Omega_2$ with the radius $r_2=0.8$ (see Fig. \ref{fig:domain_structure}a) and a spherical layer representing the outer sourceless area $\Omega_1$ with the radius $r_1=1$. The conductivity of the inner domain is $\sigma(\Omega_2) = \sigma_2 =2.2  S/m$, and the domain $\Omega_1$ can be characterized with the conductivity $\sigma(\Omega_1) = \sigma_1 = 0.1 S/m$. The electrical sources are depicted with the support (three spots) on Fig.\ref{fig:results_1} a). The support spots contain the constant currents inside it: $\JJ(\xx) = (0, 10^{-7}, 0), \xx\in supp(\JJ)$ A. 

Fig.\ref{fig:results_1}a) and b) depict the simulated electric potential on the outer surface and the surface of the "active" domain, respectively, while c) and d) show respectively the simulated potential on the 'brain' surface, and the result of its reconstruction using the algorithm presented above.  

The finite element mesh for the sourceless domain $\Omega_1$ consists of $484946$ tetrahedra and $90435$ vertices. The calculation time for the Cauchy problem solver implemented with the Matlab is equal to $5.7$ seconds with a laptop equipped with $16$ Gb RAM and CPU Intel Core-i7 7820HQ.

\begin{figure}
	\begin{tabular}{cc}
		\includegraphics[width=0.45\linewidth]{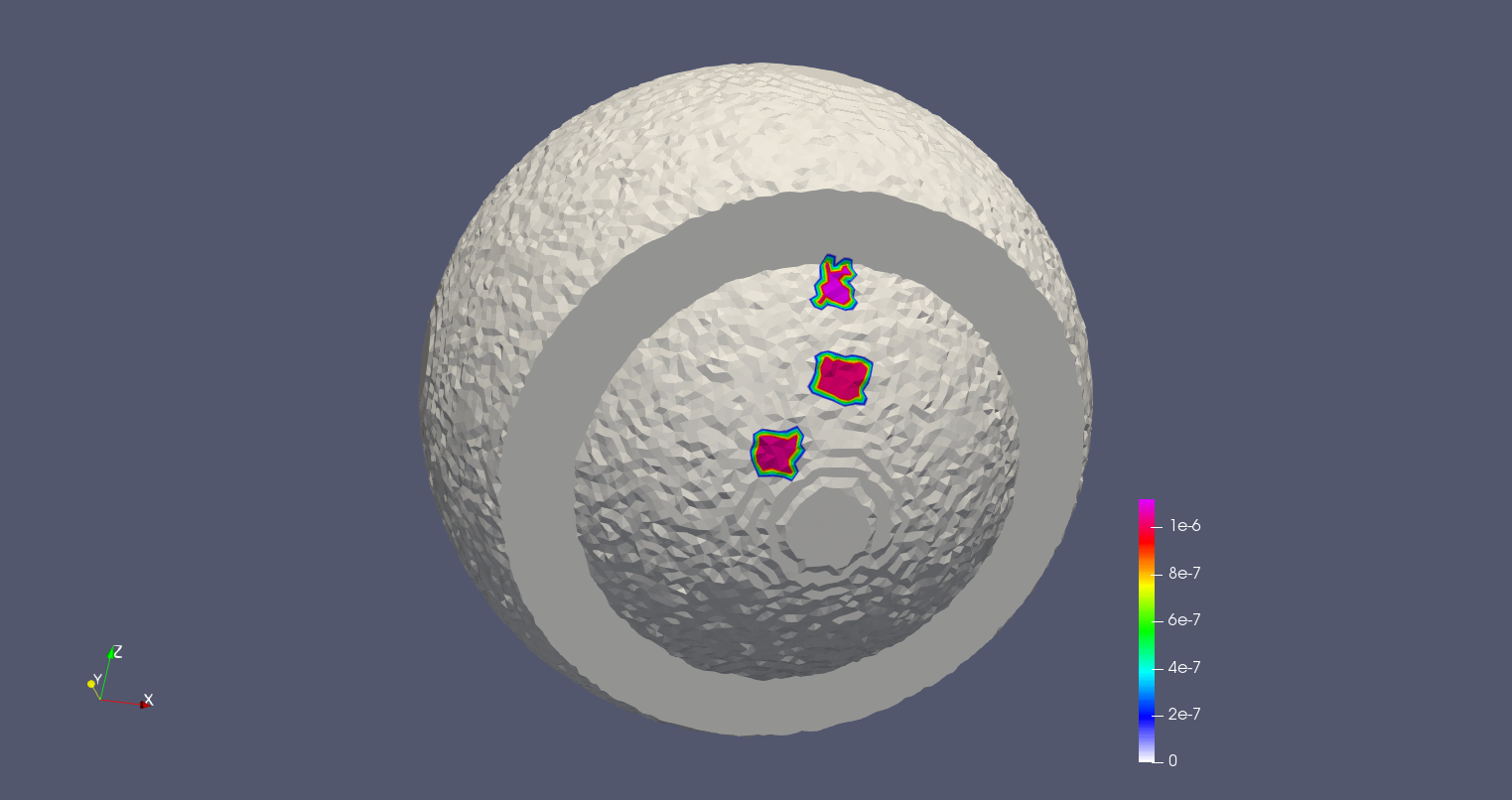} & 
		\includegraphics[width=0.45\linewidth]{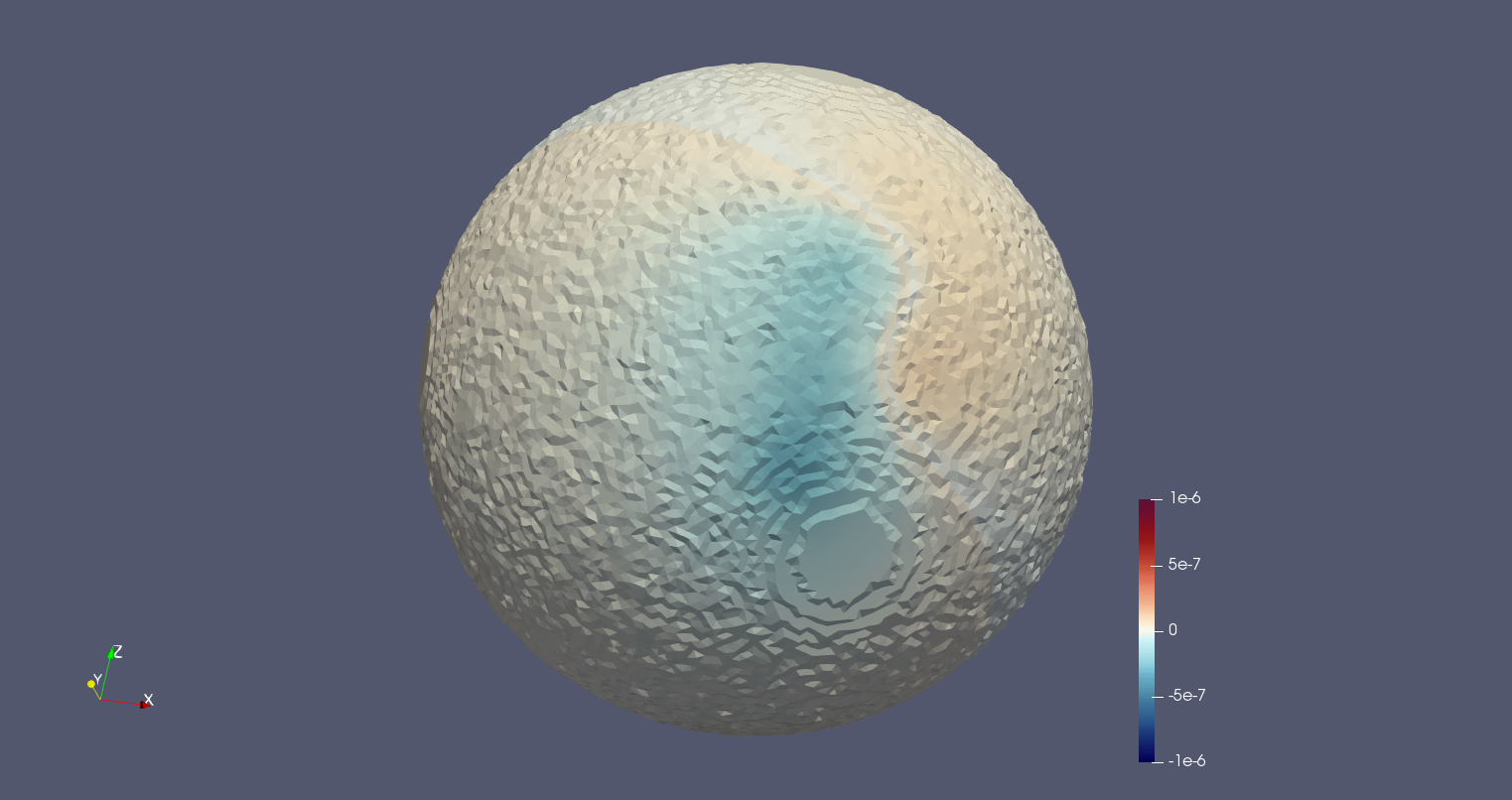} \\ 
		(a) & (b) \\
		\includegraphics[width=0.45\linewidth]{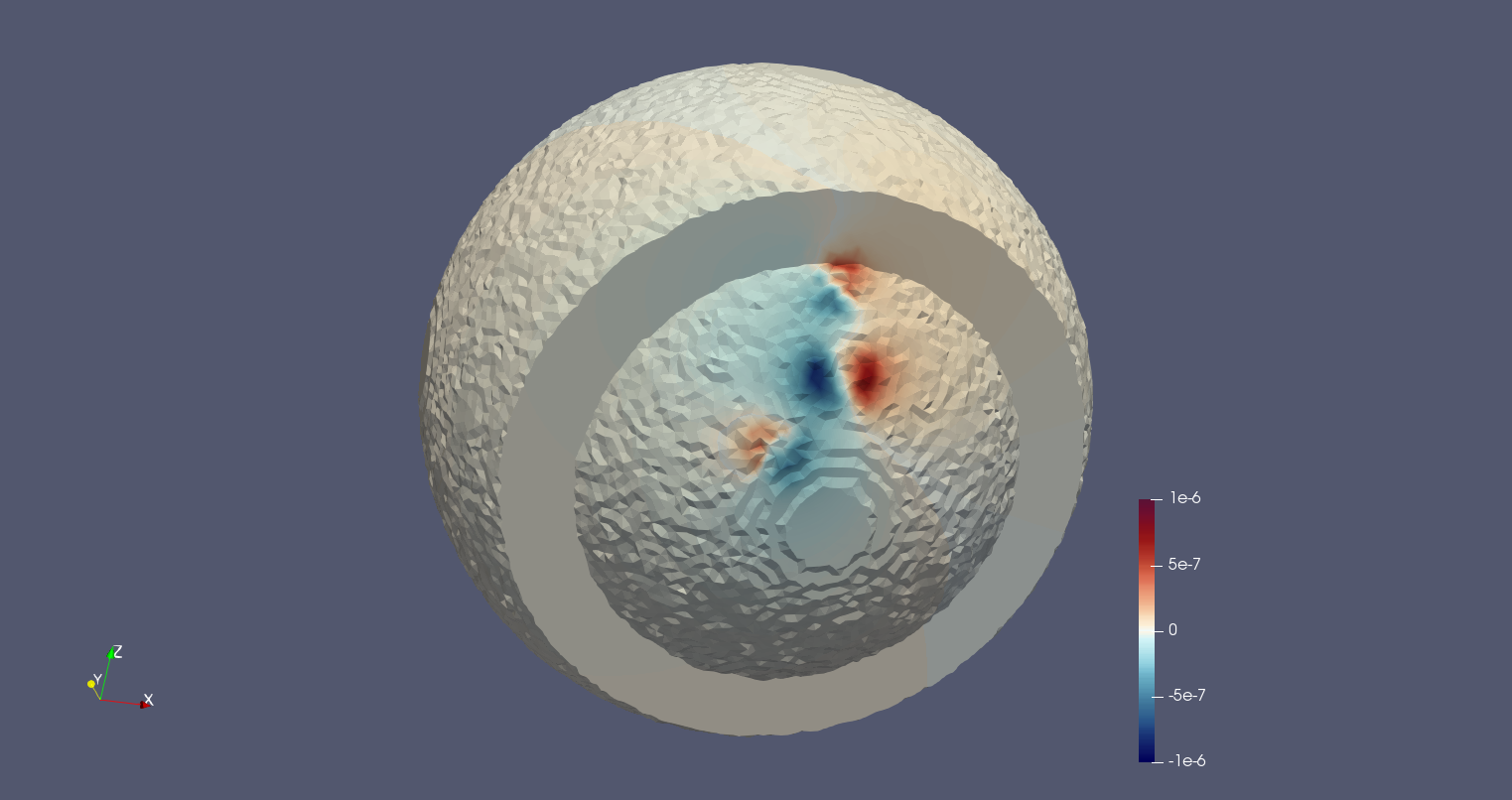} & 
		\includegraphics[width=0.45\linewidth]{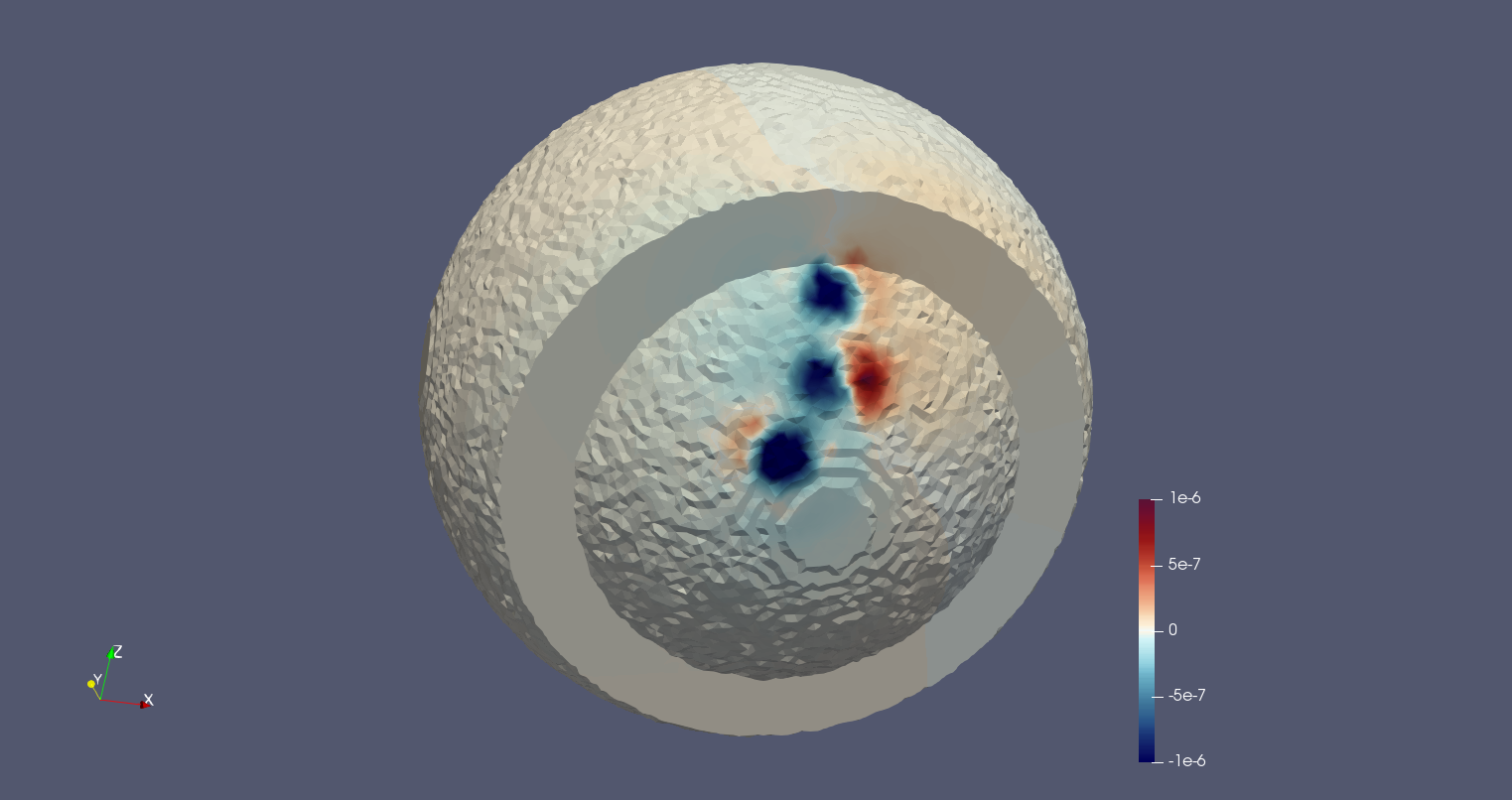} \\
		(c) & (d)
	\end{tabular}
	\caption{Spherical model (sources located on the surface of the inner sphere): a) The current density (sources); b) simulation of the electric potential on the outer surface (scalp); b) simulation of the electric potential on the "active" domain surface; c) the potential on the inner sphere surface, obtain vial solution of the Cauchy problem with regularization parameters $\epsilon=0.00001$ and $\delta=0.001$
		}
	\label{fig:results_2}
\end{figure}

\subsection{Spherical model: more realistic case}
Geometrically, the model under consideration in the current subsection is identical to the one in the previous subsection. It also characterized with the same conductivity distribution. The difference is in the current. First of all, the currents here are located on the interface between active and inactive domains, which is closer to the real-life encephalographic situation. Strictly speaking, the domain $\Omega_1$, in this case, should not contain its $\dO_2$ boundary surface, and it should be the opened domain. However, since the potential can be calculated via the solution of the Cauchy problem for the surface, which is arbitrary close to the surface $\dO_2$, we were not surprised that the modelled and restored via the Cauchy problem solution potentials on an active domain surface are relatively close to each other. The following results can be seen in Fig.\ref{fig:results_2}.

The finite element mesh for the sourceless domain $\Omega_1$ consists of $484946$ tetrahedra and $90435$ vertices. The calculation time for the Cauchy problem solver implemented with the Matlab is equal to $5.7$ seconds with a laptop equipped with $16$ Gb RAM and CPU Intel Core-i7 7820HQ.

\begin{figure}
	\begin{tabular}{cc}
		\includegraphics[trim=360 0 250 0,clip,width=0.45\linewidth]{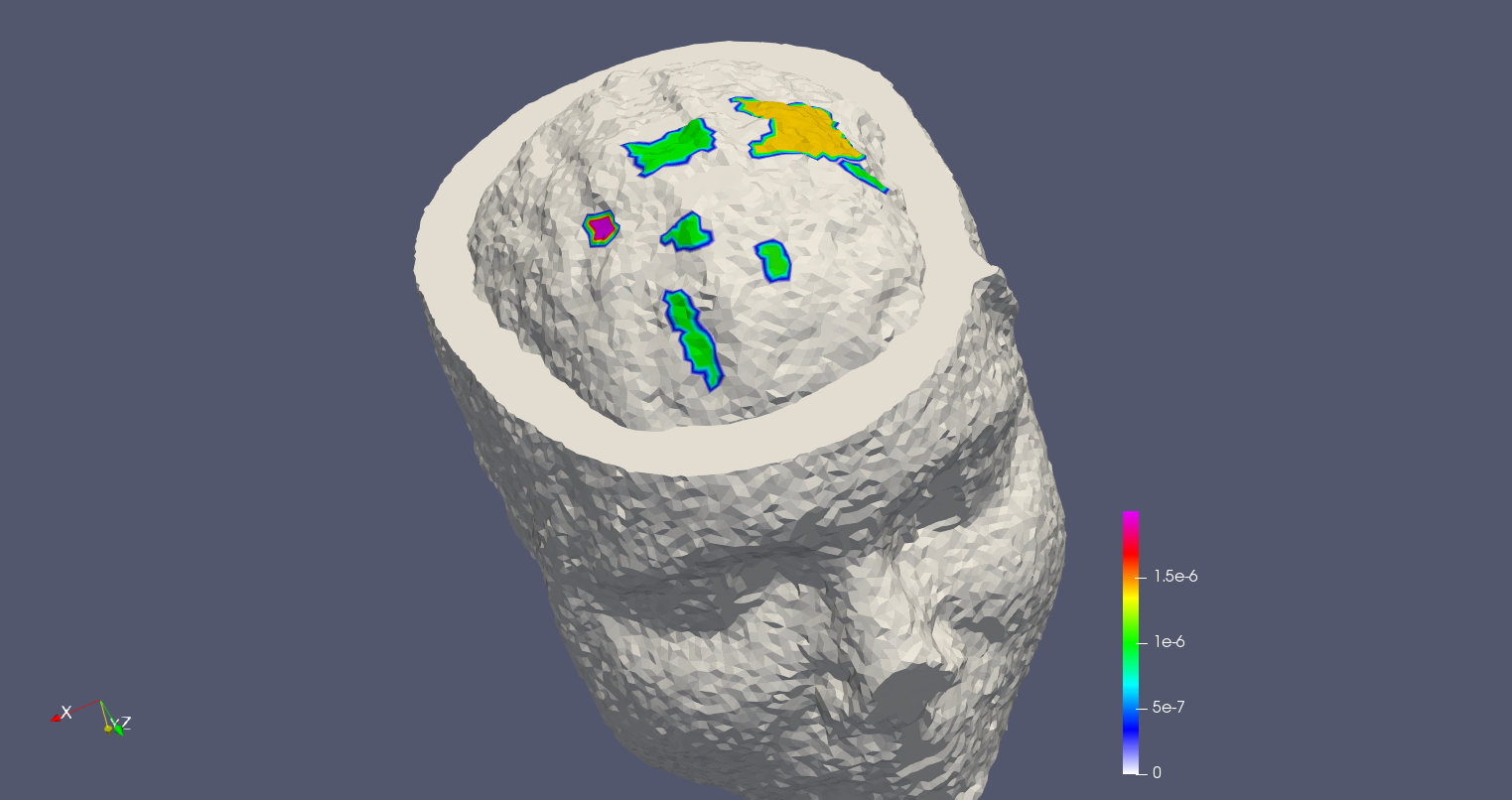} & 
		\includegraphics[trim=360 0 250 0   ,clip,width=0.45\linewidth]{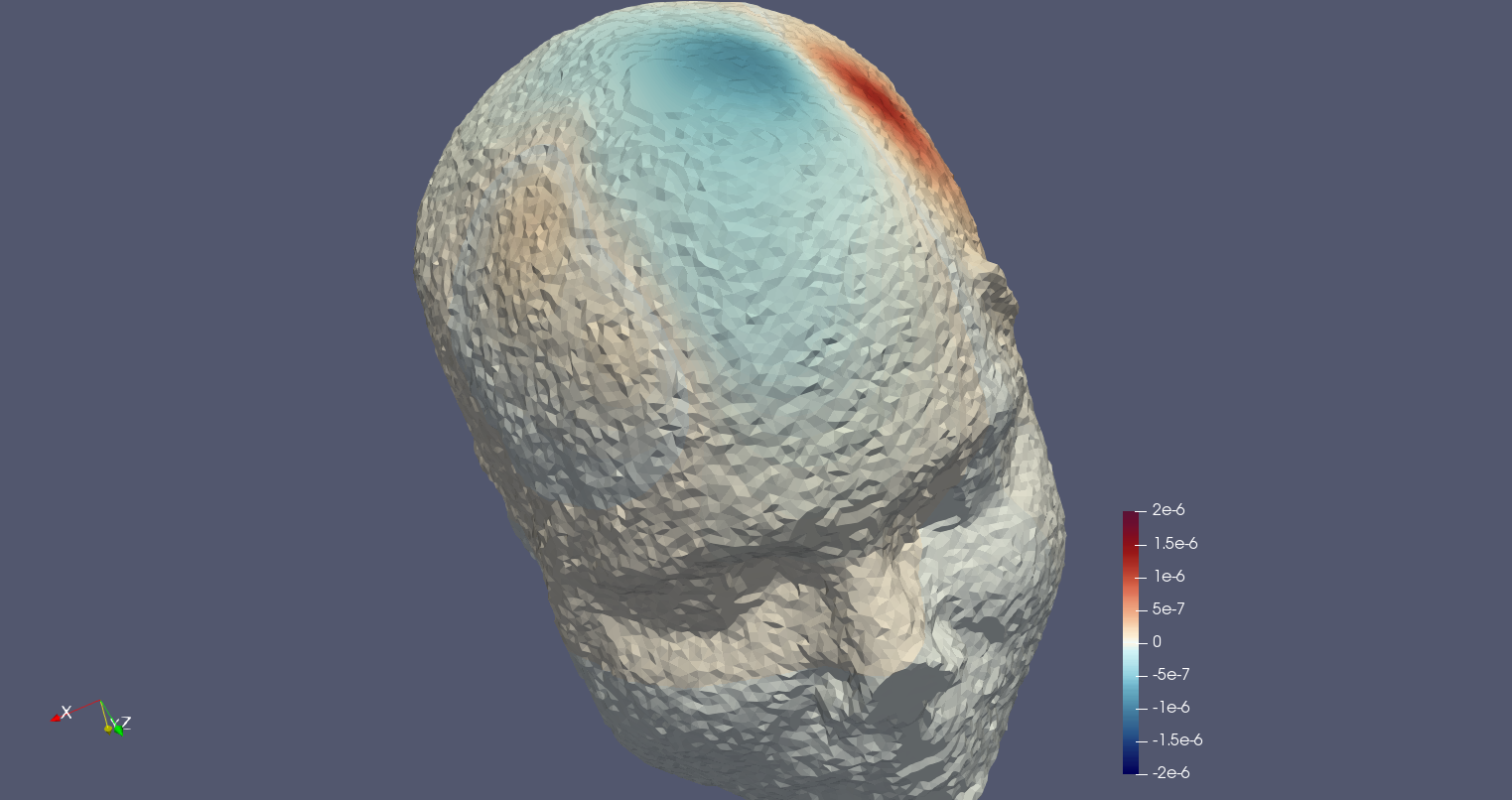} \\
		(a) & (b) \\
		\includegraphics[trim=360 0 250 0	,clip,width=0.45\linewidth]{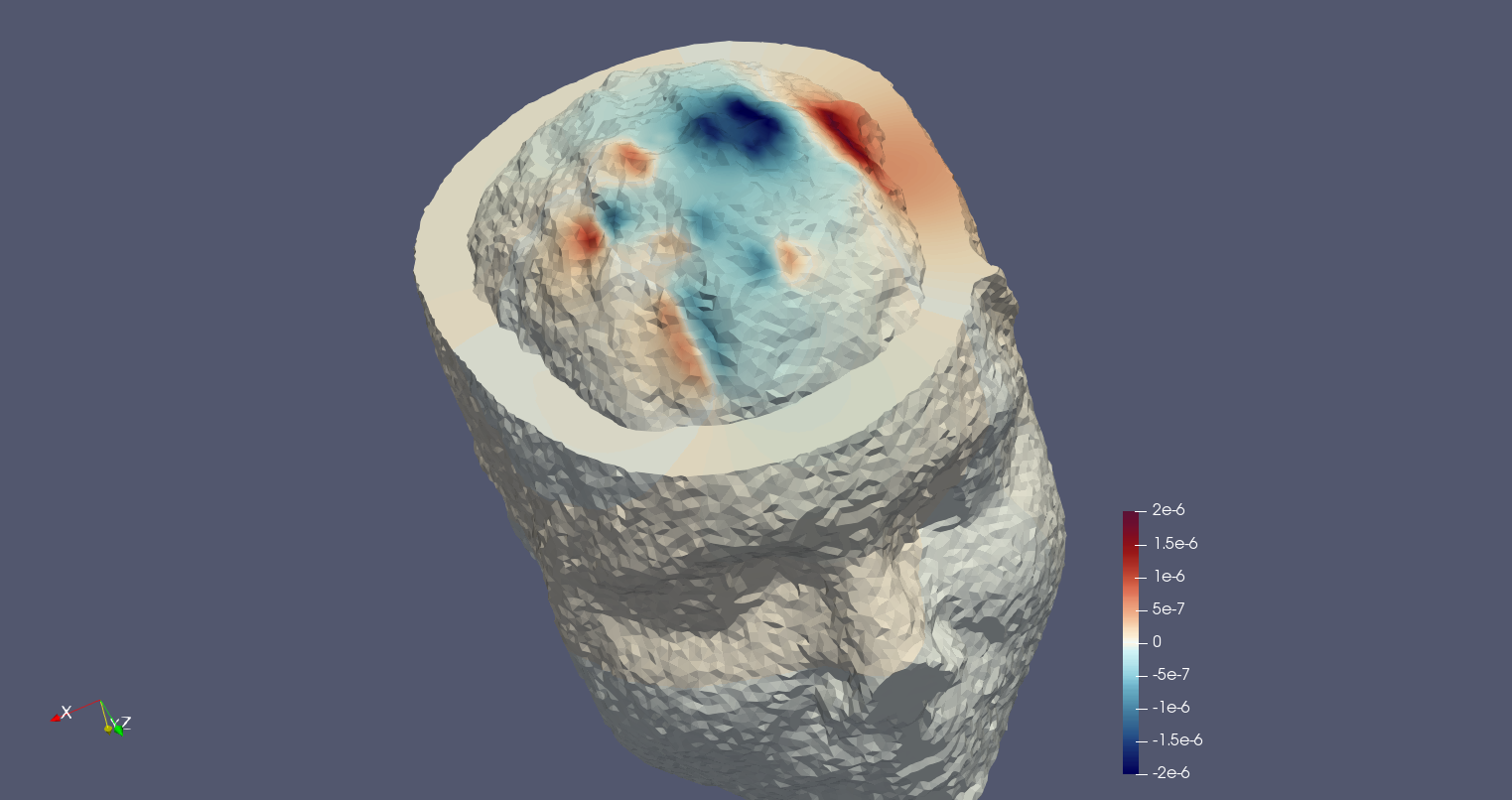} & 
		\includegraphics[trim=360 0 250 0	,clip,width=0.45\linewidth]{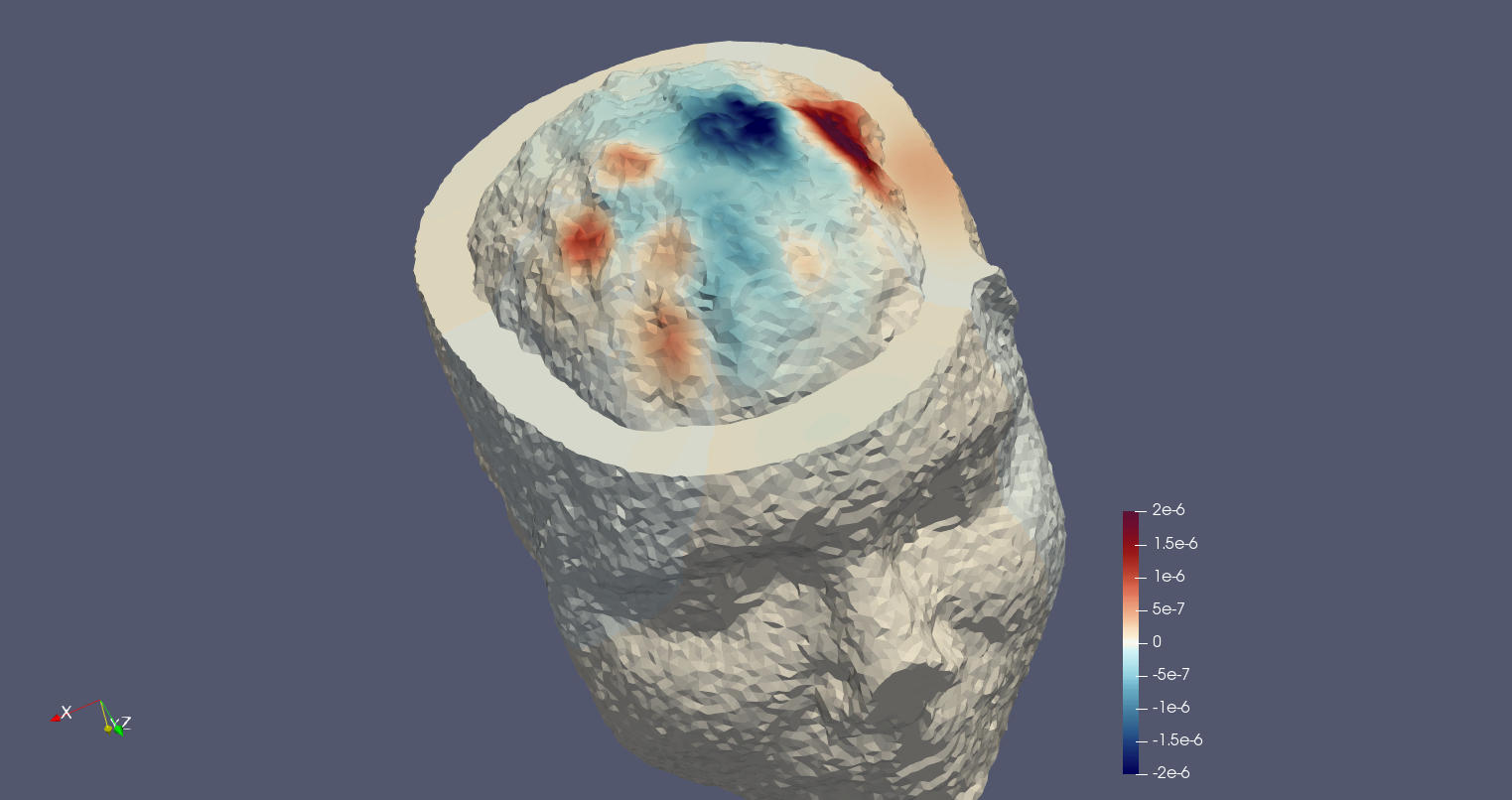} \\
		(c) & (d)
	\end{tabular}
	\caption{MRI-based model of the head: case with piecewise-constant current distribution on the brain surface. 
	(a) - the current, (b) - FEM-simulated potential on the outer surface of the head (enter data for the Cauchy problem, 
	(c) - FEM-simulated potential; (d) - reconstruction of the potential on the brain surface on the base of the 
	data, depicted with (b).}
	\label{fig:results_3}
\end{figure}

\subsection{Simplified head model based on real MRI data}
The model described in this subsection is a simplified but more realistic model of the head. The mesh was constructed using the iso2mesh software \cite{iso2mesh}. Here the "realistic" distribution of the current density over the brain surface was used. 

\begin{figure}
	\begin{center}
		\begin{tabular}{cc}
			\includegraphics[trim=360 0 250 0,clip,width=0.45\linewidth]{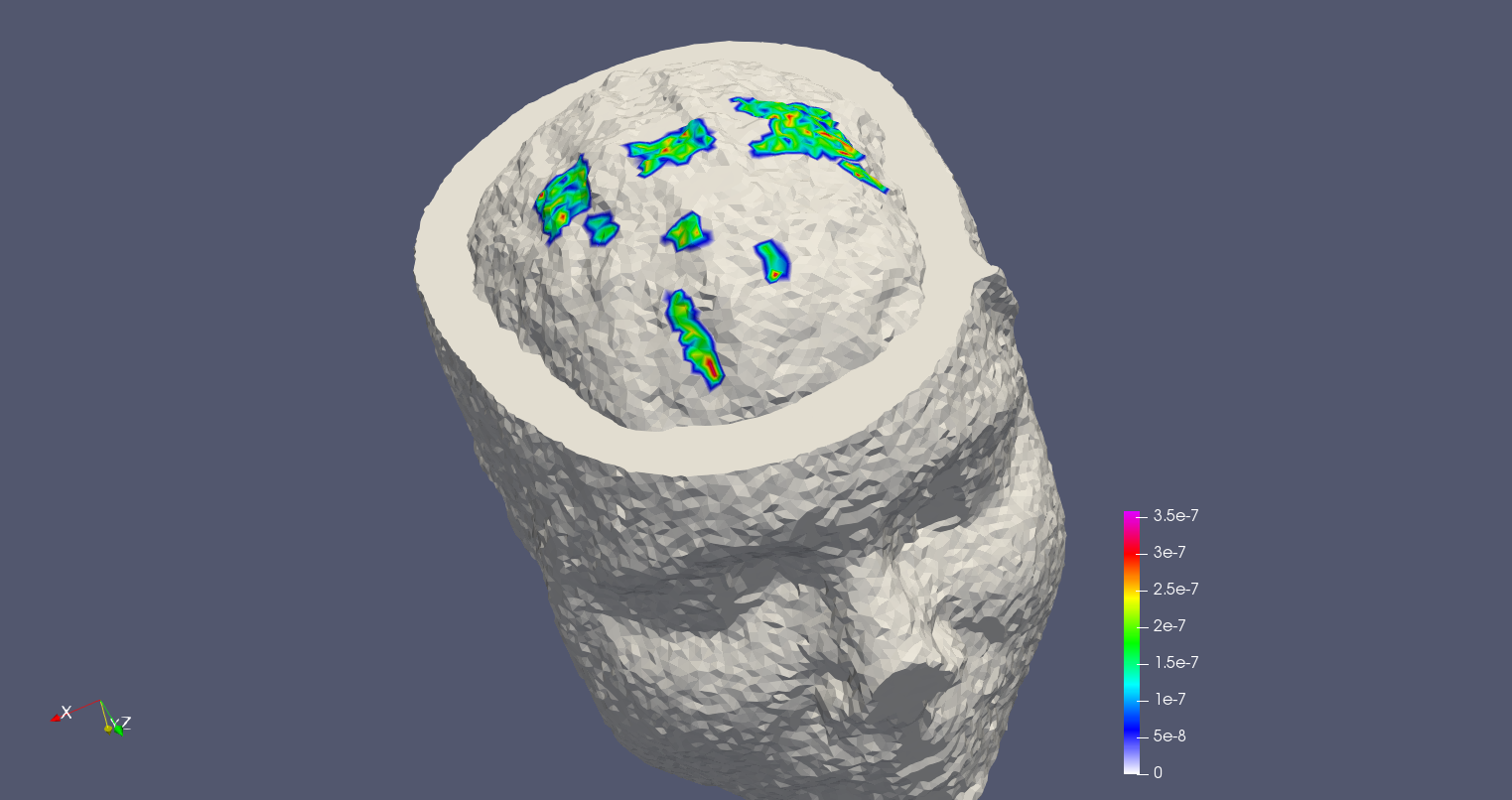} & 
			\includegraphics[trim=360 0 250 0   ,clip,width=0.45\linewidth]{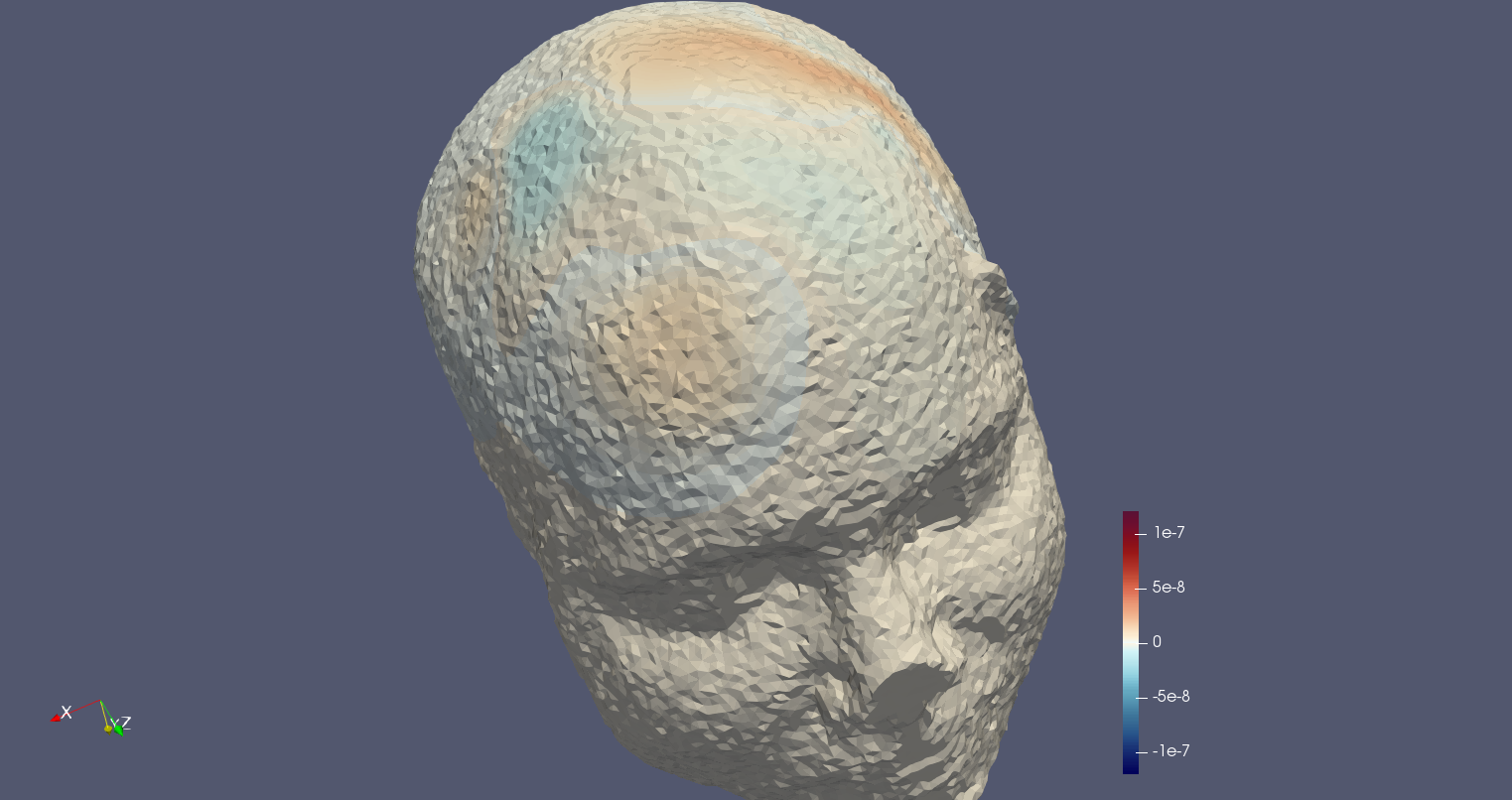} \\
			(a) & (b) \\
			\includegraphics[trim=360 0 250 0	,clip,width=0.45\linewidth]{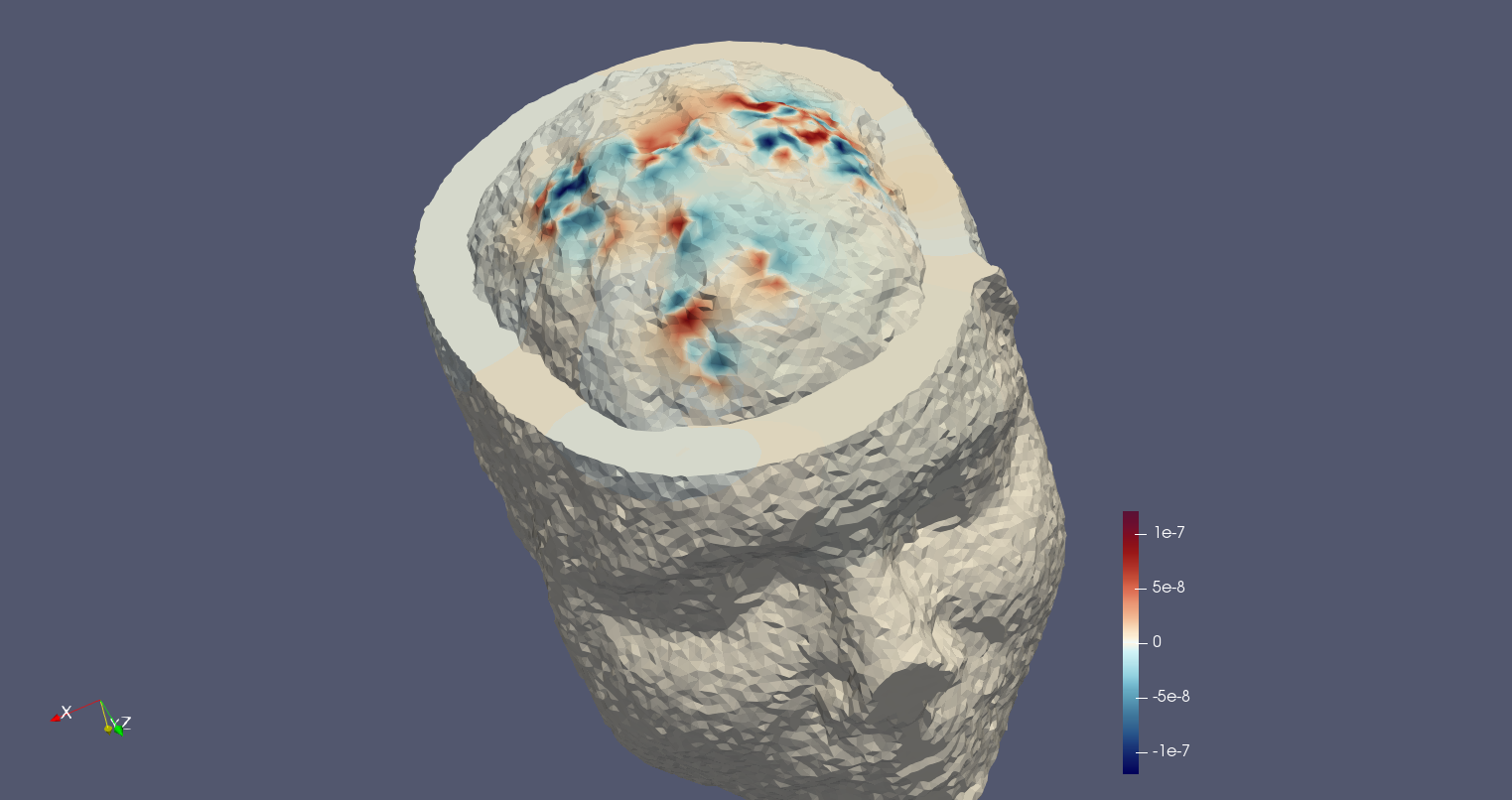} & 
			\includegraphics[trim=360 0 250 0	,clip,width=0.45\linewidth]{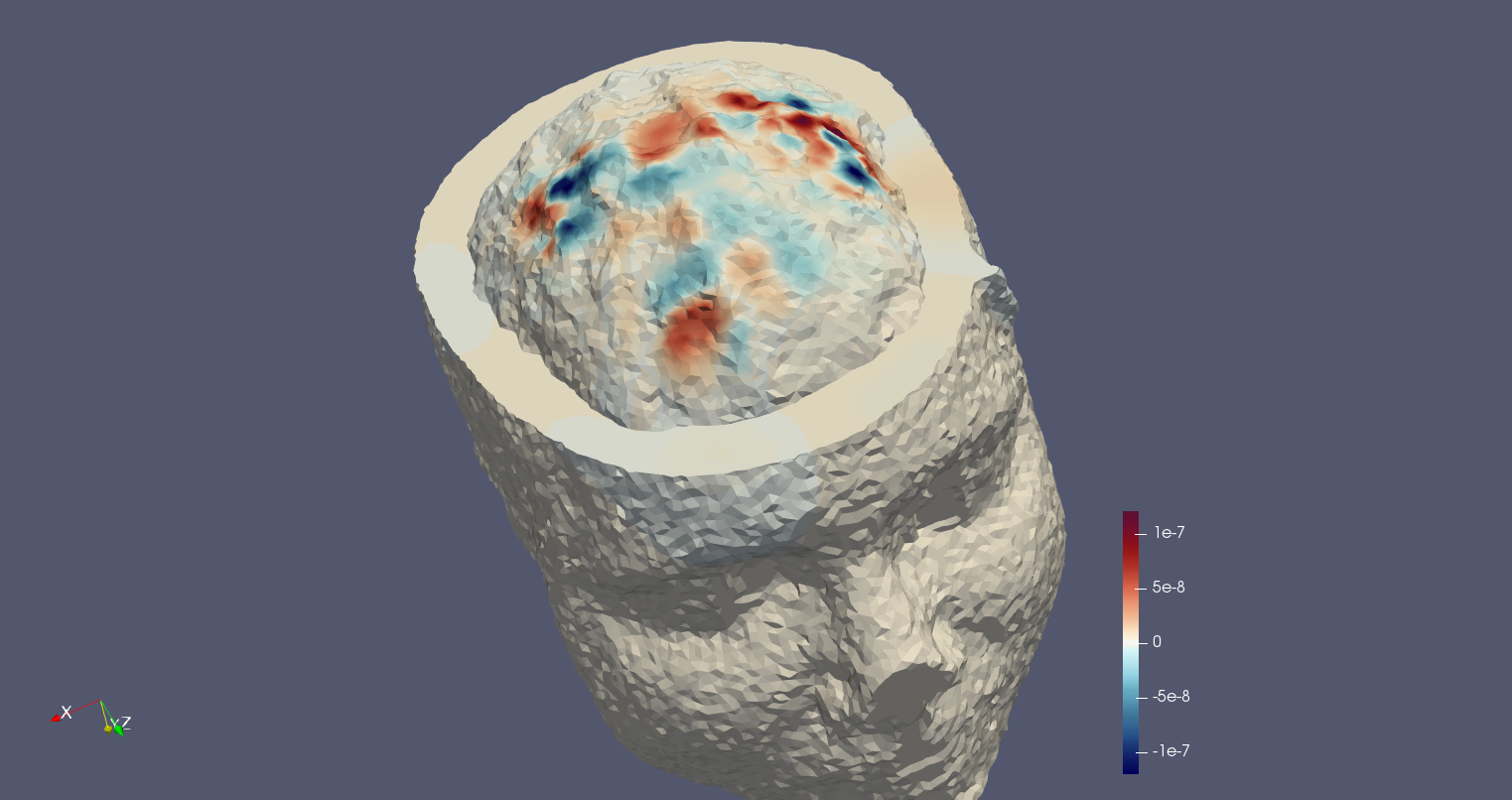} \\
			(c) & (d)
		\end{tabular}
	\end{center}
	\caption{MRI-based model of the head: case with random current, distributed within big areas of the brain surface. 
	(a) - the current, (b) - FEM-simulated potential on the outer surface of the head (enter data for the Cauchy problem, 
	(c) - FEM-simulated potential; (d) - reconstruction of the potential on the brain surface on the base of the 
	data, depicted with (b).}
	\label{fig:results_4}
\end{figure}

In order to make the result more realistic, we employed two kinds of current density distributions. Fig \ref{fig:results_3} depicts the model with several 'active' spots (support areas) of the cortex, where the current, corresponding to the spot is constant inside it. The Fig \ref{fig:results_4} shows the situation, when the current is randomely defined in each point inside the support areas (areas itself are similar to the previous case). 

The finite element mesh for the sourceless domain $\Omega_1$ consists of $444916$ tetrahedra and $82239$ vertices. The calculation time for the Cauchy problem solver implemented with the Matlab varies in the interval $6.1-11.2$ seconds in dependence of the current distribution used for modelling. All calculations have been provided with a laptop equipped with $16$ Gb RAM and CPU Intel Core-i7 7820HQ. 

\section{Conclusion}
In this paper we presented an approach for Scalp-to-Cortex data mapping, based on finite elements computational scheme. The application of the approach shows great spatial resolution of the result, which is comparable to intracranial EEG (iEEG, ECoG) techniques. The method based on solving the Cauchy problem for the Laplace's equation in homogeneous compartment and can be described as propagation of EEG data collected on the scalp through skull to the brain surface. 
Our numerical implementation employs the approximation of the first order finite elements on tetrahedral grids, which enables to model sophisticated irregularly shaped conductors with high precision. The method is considered as a pre-processing procedure, after which classical inverse source localization problem solvers can be applied.
Notably, our algorithm successfully works with minimal soft- and hardware capabilities and does not require professional knowledge. All calculations were provided in Matlab, using only a quad-core laptop and without explicit parallelization; the computational algorithm thus has a big potential on optimization. This allows us to assume that with further improvements our method can be successfully implemented for usage with mobile platforms.  
Speaking of future work, we believe the solution of the Cauchy problem described in this paper can be easily adopted for source localization in magnetoencephalography. The other line of our work is to consider the same problem but with more complex algorithms in order to achieve better accuracy and compare numerically the spatial resolution of conventional EEG, ECoG and our method. Both of these areas will be certainly reflected in our future publications.

\bibliographystyle{unsrt}
\bibliography{my}{}

\end{document}